\documentclass[12pt]{iopart}

\newcommand{\be}{\begin{eqnarray}}
\newcommand{\ee}{\end{eqnarray}}
\newcommand{\ket}[1]{\ensuremath{\left| {#1} \right>}}
\newcommand{\bra}[1]{\ensuremath{\left< {#1} \right|}}

\newcommand{\Beplus}{\ensuremath{{^9}{\rm Be}^{+} \,}}
\newcommand{\Mgplus}{\ensuremath{{^{24}}{\rm Mg}^{+} \,}}
\newcommand{\MgplusH}{\ensuremath{{^{25}}{\rm Mg}^{+} \,}}

\newcommand{\BM}{\Beplus--\,\Mgplus}
\newcommand{\MB}{\Mgplus--\,\Beplus}
\newcommand{\BMMB}{\Beplus--\,\Mgplus--\,\Mgplus--\,\Beplus}

\usepackage{graphicx}
\usepackage{iopams}	
\begin{document}

\title{Normal modes of trapped ions in the presence of anharmonic trap potentials.}

\author{J. P. Home$^{1,2}$\footnote{Author to whom any correspondence should be addressed.}, D. Hanneke$^1$, J. D. Jost$^1$, D. Leibfried$^1$, and  {D. J. Wineland}$^1$}

\address{$^1$ National Institute of Standards and Technology, 325 Broadway, Boulder, Colorado, USA}

\address{$^2$ Institute of Quantum Electronics, ETH Z{\"u}rich, Schafmattstrasse 16, 8093-Z{\"u}rich, Switzerland}

\address{E-mail: jhome@phys.ethz.ch}

\begin{abstract}
We theoretically and experimentally examine the effects of
anharmonic terms in the trapping potential for linear chains of
trapped ions. We concentrate on two different effects that become
significant at different levels of anharmonicity. The first is a
modification of the oscillation frequencies and amplitudes of the ions' normal
modes of vibration for multi-ion crystals, resulting from each ion
experiencing a different curvature in the potential. In the second
effect, which occurs with increased anharmonicity or higher excitation amplitude, amplitude-dependent shifts of the normal-mode frequencies become important. We
evaluate normal-mode frequency and amplitude shifts, and comment on the
implications for quantum information processing and quantum state
engineering. Since the ratio of the anharmonic to harmonic terms
typically increases as the ion--electrode distance decreases,
anharmonic effects will become more significant as ion trap sizes
are reduced. To avoid unwanted problems, anharmonicities should therefore be
taken into account at the design stage of trap development.
\end{abstract}

\maketitle

\section{Introduction}
Quantum control of the motional degrees of freedom of trapped atomic
ions is among the most advanced in physics. It has an important role
in realizing deterministic quantum information processing with trapped
ions and offers a rich playground for the exploration of quantum-state
engineering and quantum control. Examples of the latter include the
production of Fock states, squeezed states, coherent states, and superpositions
of coherent states analogous to the {}``Schr\"{o}dinger's
cat'' thought experiment \cite{96Meekhof,96Monroe,04Haljan,07McDonnell,poschingerPRL2010}.
The motional states of trapped ions have been a key ingredient
for simulation of quantum-optical systems \cite{leibfriedPRL2002} and the Dirac equation \cite{10Gerritsma,gerritsmaPRL2011} and for the realization
of quantum walks in phase space \cite{09Schmitz,10Zahringer}.

For ion-trap quantum information processing~\cite{08Blatt}, the collective motion
of the ions is critical for performing deterministic multi-qubit quantum
gates. Recent work includes the entanglement of up to 14 ions \cite{10Monz},
and a two-qubit entangled state fidelity of $99.3\,\%$~\cite{08Benhelm}. In
order to perform more complex operations, higher precision control
of both the internal and motional states of the ions will be required,
and scalable techniques will need to be implemented \cite{09Home}.
The high operation fidelities of around $99.99\,\%$ required to achieve
fault-tolerance for quantum information processing \cite{03Steane,05Knill}
place stringent demands on the motional control.

Most of the experimental work performed to date has used traps
where the
potential is harmonic to a very high degree. The resultant normal-mode frequencies and amplitudes have been extensively studied \cite{97Steane2,98James1,01Morigi}.
The relative importance of anharmonic terms depends primarily on their length scales relative to those relevant to the ions. In this paper, we consider two important effects that occur at different length scales. First, weak anharmonicity modifies the normal modes of multi-ion chains when anharmonic terms become significant over the length $L$ of a chain of ions. In this case the ions' equilibrium positions are modified and the local curvature
of the potential is different at the position of each ion. This modifies both the
frequency and ion amplitudes of the normal modes of the chain, but
the normal-mode eigenstates are still stationary states of the Hamiltonian. Second, strong anharmonicity plays a role when changes in curvature over the size of the motional wavefunction become significant.
This creates motional frequency shifts as a function of the motional
excitation and cross-coupling between different normal modes. The effects of strong anharmonicity have been observed when
caused by anharmonicity in the Coulomb interaction between the ions \cite{08Roos,09Nie}; we extend these results to include anharmonicity in the trapping potential. The Coulomb anharmonicity plays a crucial role in the prediction of a temperature-driven structural phase transition in ion chains~\cite{08Fishman,gongPRL2010}, and anharmonic trap potentials will modify the effect and perhaps allow it to be tuned.
Although we focus in this paper on anharmonicity as a perturbation to harmonic trap potentials, several experiments use strongly anharmonic potentials to split and recombine chains of ions \cite{09Home,02Rowe,09Jost,10Hanneke} and to engineer ion chains with nearly equal ion spacing \cite{linEPL2009}.

Manipulating ions in small-scale traps
that can be microfabricated is useful for large-scale quantum information processing \cite{09Home,07Steane,10Amini}; for achieving more intricate motional
control \cite{09Jost,04Porras,08Deng,brownNature2011,harlanderNature2011}; and for coupling
trapped ions to other quantum devices \cite{04Tian,98Wineland2,90Heinzen,05Hensinger}. Anharmonic terms have the potential to scale unfavorably as trap sizes are reduced. Consider a trap with some characteristic length scale $\rho$. This length might be, for example, the distance from a trapped ion to the nearest point on the surface of a trap electrode. In an expansion of the trap potential $V_t$, the size of the $n$th term is proportional to $\partial^n V_t/\partial z^n$ and thus it will tend to scale as $\rho^{-n+2}$ relative to the harmonic term. Moreover, some recent microfabricated traps employ a geometry in which the ion resides above a surface containing the electrodes \cite{10Amini,06Seidelin,leibrandtQIC2009,daniilidisNJP2011}. The intrinsic asymmetry in these surface-electrode traps can lead to large odd-order anharmonicities in the direction perpendicular to the electrode surface. While anharmonicities should tend to increase with decreasing trap size, the trap potential can still be engineered to suppress them, provided the trap is designed with sufficient degrees of freedom.

To give a sense of the typical lengths involved, we note that a single \Beplus ion in the ground state of a $1$~MHz harmonic potential has a root-mean-square wavefunction size of $\sigma=24$~nm. Two ions in the same potential are separated by $L=9$~$\mu$m. Recent trends in quantum information are towards increased numbers of ions to perform multi-qubit operations simultaneously on many qubits \cite{10Monz,05Haffner,05Leibfried,10Barreiro,11Islam},
to accommodate sympathetic refrigerator
ions along with those used to store qubits \cite{09Home,09Jost,10Hanneke,linEPL2009},
and to encode logical qubits in multiple
physical qubits \cite{kielpinskiScience2001,05Langer,09Monz}. For chains containing eight ions, trap frequencies along the chain direction between 1~MHz and 5~MHz lead to a range of $L$ between 36~$\mu$m and 12~$\mu$m.

The paper is organized as follows. After a brief introduction to calculation
methods for normal modes of trapped-ion crystals,
we discuss weak anharmonicities, giving simple examples, experimental
results, and methods for characterization. Next, we
theoretically treat effects arising from stronger anharmonicities.
We evaluate these effects with reference to a surface-electrode trap that has been used at NIST. In both sections,
we discuss the implications of our results for precise control of
trapped-ion chains for quantum information and state-engineering,
focusing on multi-qubit quantum logic gates that use the ions' normal modes. Finally, we provide a short discussion of
the susceptibility of anharmonic traps to electric field drift, after
which we conclude.

\section{Calculating normal modes of trapped ion chains} \label{sec:calc_normal_modes}

Normal-mode frequencies and amplitudes may be calculated from the classical Lagrangian equations of motion and quantized with the usual harmonic-oscillator algebraic formalism.
We consider the case of $N$ ions each with charge $q$ in a potential
well. The energy of the ions is given as a sum of kinetic- and potential-energy terms $E=T+U$, where
\begin{eqnarray}
	T & = & \sum_{i=1}^{N}\frac{m_{i}}{2}\dot{\bi{z}}_{i}^{2},\nonumber \\
	U & = & \sum_{i=1}^{N}q V_{t}(\bi{z}_{i},m_i)+\frac{1}{2}\sum_{{j,i=1\atop j\neq i}}^{N}\frac{q^{2}}{4\pi\epsilon_{0}|\bi{z}_{i}-\bi{z}_{j}|}\ \ \ .
\end{eqnarray}
Here $m_{i},\bi{z}_{i}$ denote the mass and position of the $i$th
ion. The trap potential $V_t$ includes both a mass-independent static potential and a mass-dependent pseudopotential  arising from the radiofrequency confinement \cite{98Wineland2}. In what follows, it is convenient to write the coordinates as $3N$ scalar parameters $z_{1},\ldots,z_{3N}$ that produce
the $N$ vectors $\bi{z}_{1},\ldots,\bi{z}_{N}$.

The set of equations $\partial U/\partial z_{i}=0$ give the set of equilibrium positions $\{z_{i}^{0}\}$ for the
ions.
For large $N$ this is done by numerical minimization, since the analytical
expressions become complicated. In a Taylor expansion of the potential around
these equilibrium positions, the leading term is
at second order, which gives the symmetric Hessian matrix
\begin{equation}
	 H_{ij}^{\prime}=\left.\frac{1}{\sqrt{m_{i}m_{j}}}\frac{\partial^{2}U}{\partial z_{i}\partial z_{j}}\right|_{\{z_{i}^{0}\}} .
	\label{eq:hessian}
\end{equation}
As is typical \cite{LLmechanics}, we use mass-weighted coordinates; we indicate this transformation with a prime, $z_{i}^{\prime}=\sqrt{m_{i}}z_{i}$. This transformation allows us to write
the kinetic energy $T$ in a form that is independent of mass.

The normal modes and their
corresponding frequencies can be found by solving the Lagrangian
equations of motion for the system. In this case, the relevant quantities are the displacements
from equilibrium, $\zeta_{i}^\prime=z_i^\prime-z_{i}^{0\prime}$. Neglecting higher orders in $U$ than those described by \eref{eq:hessian}, the $3N$ equations of motion are
\begin{equation}
	 \ddot{\zeta}_{j}^{\prime}+\sum_{i=1}^{3N}H_{ij}^{\prime}\zeta_{i}^{\prime}=0 .
	\label{eq:vector_transformation_into_normal_mode_basis}
\end{equation}
Inserting a fiducial solution $\zeta_j^\prime = \zeta_j^{0\prime}e^{i\omega t}$ gives a linear system of equations that can be diagonalized to yield
the normal modes of the system, which are defined by the eigenvalues
and eigenvectors of the matrix $H_{ij}^{\prime}$. The eigenvalues
are equal to $\omega_{\alpha}^{2}$, where $\omega_{\alpha}$ is the
motional frequency of the normal mode $\alpha$. The matrix of eigenvectors
$e_{i}^{\prime\alpha}$ allows us to express normal-mode coordinates as
a function of the individual ion coordinates by use of
\begin{equation}
	 \zeta_{\alpha}^{\prime}=\sum_{i=1}^{3N}e_{i}^{\prime\alpha}\zeta_{i}^{\prime}.
\end{equation}
 Since each normal mode acts as an independent oscillator, we can
quantize them in the usual manner, writing the mass-weighted position operator as
\begin{equation}
	 \hat{\zeta}_{\alpha}^{\prime}=\sigma_{\alpha}^{\prime}\left(\hat{a}_{\alpha}+\hat{a}_{\alpha}^{\dagger}\right) ,
	\label{eq:quantized_position_operator}
\end{equation}
 where $\sigma_{\alpha}^{\prime} = \sqrt{\hbar/(2\omega_{\alpha})}$ and $\hat{a}_\alpha^\dagger$, $\hat{a}_\alpha$ are the raising and lowering ladder operators. For the $i$th ion in mode $\alpha$, the ground-state wavefunction root-mean-square size is
\begin{equation}
	\sigma_i = \frac{1}{\sqrt{m_i}} e_i^{\prime\alpha} \sigma_\alpha^\prime .
	\label{eq:rmsWavefunction}
\end{equation}
The quantized form for the ion's excursion from equilibrium is \cite{01Morigi}
\begin{equation}
\hat{\zeta}_{i}=\frac{1}{\sqrt{m_{i}}}\sum_{\alpha=1}^{3N}(e_{i}^{\prime\alpha})^{-1}\sigma_{\alpha}^{\prime}\left(\hat{a}_{\alpha}+\hat{a}_{\alpha}^{\dagger}\right) .
\end{equation}

\subsection{One-dimensional simplification}

For much of the paper, and in many experiments, we are primarily interested
in the dynamics of motion along
the ion chain. For an axial trap potential centred at $z=0$, we can expand it as a power series:
\begin{equation}
	V_{t}(z) 	= \sum_{n=2}^{\infty}\kappa_{n}z^{n}
						= \kappa_{2}z^{2}\left[1+\sum_{n=3}^{\infty}\left(\frac{z}{\lambda_n}\right)^{n-2}\right] ,
\end{equation}
 where $\lambda_{n} = (\kappa_{n}/\kappa_{2})^{1/(2-n)}$ is a length used to parameterize the anharmonicity.
We will consider potentials where the harmonic ($n=2$) term dominates, as is typically the case in the experiments described below. In terms of the equation above, we assume $|\Lambda/\lambda_n|^{n-2}\ll1$, where $\Lambda$ is some relevant length scale such as the ion chain length $L$ or the ground-state wavepacket size $\sigma$. We assume no axial pseudopotential, so the series co-efficients do not depend on the ion mass.


\section{Anharmonic modifications to normal modes}

For anharmonic perturbations that appear on length scales comparable to that of the ion chain but still negligible on the scale of an ion wave-packet, the main effects are modifications of the normal-mode frequencies and amplitudes. For pairs of ions, these may be examined analytically, and we give examples of cubic and quartic potentials for pairs with equal and with unequal masses. The shifts seen are similar to those in longer chains, which are more easily analyzed numerically. We then examine these effects experimentally. We measure the anharmonic frequency shifts in chains of up to eight ions of equal mass. With mixed-species ion chains, we measure modified mode amplitudes and demonstrate a technique for nulling odd-order anharmonicities.

\subsection{\label{Illustrative-examples}Illustrative examples}

We start by
giving two simple examples with
equal-mass ions, which serve to illustrate the main effects that
are observed in larger crystals, where numerical methods are more
convenient. The parameter $l=\left(q/8\pi\epsilon_{o}\kappa_{2}\right)^{1/3}$ is a characteristic
length scale for the few-ion cases. For example, the distance between two ions in a harmonic well is $L=2^{1/3}l$. The length of a chain of $N$ ions in a harmonic potential scales as $L\sim l N^{0.37}$, obtained from a fit to numerically calculated chain lengths. In the equal-mass examples below, both cubic and quartic perturbations create a frequency shift that scales as $\left(l/\lambda_n\right)^{2}$. We consider the same perturbations for two unequal mass ions and show that now the frequency shift scales differently in the sizes of the perturbations, $\left(l/\lambda_n\right)^{n-2}$. In addition, the cubic perturbation depends on ion-order, while the quartic one does not, allowing the two types of anharmonicity to be differentiated.

\subsubsection{\label{sub:Illustrative-example-1:}Example 1: cubic
term with two ions of equal mass.}

Consider a potential of the form $V_{t}(z)=(\kappa_{2}z^{2})\left(1+ z/\lambda_3\right)$, and assume that $|l/\lambda_3|\ll1$,
i.e., that the cubic term is small compared to the quadratic term over the length scale of the ion separation. The equilibrium positions $z_{\pm}^{0}$
 are found by solving the set of equations $\partial U/\partial z_{i}=0$, which has solutions (to second order
in $l/\lambda_3$)
\begin{equation}
	z_{\pm}^{0} \simeq \pm \frac{l}{2^{2/3}}\left[1\mp \frac{3}{2^{5/3}}\frac{l}{\lambda_3} + \frac{3}{2^{7/3}}\left(\frac{l}{\lambda_3}\right)^2\right].
	\label{eq:cubic_z0}
\end{equation}
Expanding the potential about these equilibrium positions,
we find that the eigenfrequencies are
\begin{eqnarray}
	\omega_{c} \simeq \sqrt{\frac{2q\kappa_{2}}{m}} \left[1- \frac{9}{2^{7/3}} \left(\frac{l}{\lambda_3}\right)^2\right] \\
	\omega_{s} \simeq \sqrt{\frac{6q\kappa_{2}}{m}} \left[1- \frac{3}{2^{7/3}} \left(\frac{l}{\lambda_3}\right)^2\right] .
\end{eqnarray}
The corresponding eigenvectors are
\begin{equation}
	e_{c}^{\prime} \simeq \frac{1}{\sqrt{2}}
	\left(\begin{array}{c}
		 1-\frac{3}{2^{5/3}}\frac{l}{\lambda_3}\\
		 1+\frac{3}{2^{5/3}}\frac{l}{\lambda_3}
	\end{array}\right)
\end{equation}
and
\begin{equation}
	e_{s}^{\prime} \simeq \frac{1}{\sqrt{2}}
	\left(\begin{array}{c}
		 -1-\frac{3}{2^{5/3}}\frac{l}{\lambda_3}\\
		 1-\frac{3}{2^{5/3}}\frac{l}{\lambda_3}
	\end{array}\right).
\end{equation}
In the limit where
$l/\lambda_3\rightarrow0$ these normal modes would correspond to the centre-of-mass
and stretch modes. We see from these results that the corrections to the eigenvectors
enter at lower order in $l/\lambda_3$ than the corrections to the eigenfrequencies,
as would be expected from general considerations of perturbation
theory \cite{BkShankar}.

\subsubsection{Example 2: quartic term with two ions of equal mass.\label{sub:Illustrative-example-2:}}

The quartic term in the trapping potential has been studied for two ions in the context of separation of ions from a single-well
potential into a double-well \cite{04Home}, in which the quartic
term was as strong as the harmonic term. Here we take an approach similar
to that of the previous section and assume it to be small compared to the
harmonic term. The potential is then of the form $V_{t}(z)=\kappa_{2}z^{2}\left[1+(z/\lambda_4)^{2}\right]$,
with $|l/\lambda_4|^{2}\ll1$. In this case the centre of the ion pair remains at $z=0$, and the equilibrium positions of the two ions
are
\begin{equation}
	z_\pm^0 \simeq \pm \frac{l}{2^{2/3}}\left[1 - \frac{1}{3\cdot 2^{1/3}}\left(\frac{l}{\lambda_4}\right)^2 + \frac{2^{4/3}}{9}\left(\frac{l}{\lambda_4}\right)^4\right].
\end{equation}
The normal mode
frequencies are
\begin{eqnarray}
	\omega_{c} \simeq \sqrt{\frac{2q\kappa_{2}}{m}}\left[1+\frac{3}{2^{4/3}}\left(\frac{l}{\lambda_4}\right)^2\right] \\
	\omega_{s} \simeq \sqrt{\frac{6q\kappa_{2}}{m}}\left[1+\frac{5}{3\cdot 2^{4/3}}\left(\frac{l}{\lambda_4}\right)^2\right].
\end{eqnarray}
Note that the frequency shift comes in as $(l/\lambda_n)^2$ for both the quartic and the cubic perturbations.
For two ions, the normal-mode amplitudes remain the centre-of-mass and stretch modes, as is the
case for a purely harmonic potential. For more than two ions, adjacent ions no longer see identical potentials, and the normal-mode amplitudes will shift from their harmonic values.

\subsubsection{Example 3: cubic term with two ions of unequal mass.}

We relax the equal mass assumption of the prior two cases, and re-examine the cubic term for ions with a mass ratio $\mu = m_1/m_2$ (with $z_1 < z_2$). Because of our assumption of no axial pseudopotential, the only mass-dependence is in the kinetic energy, so the ion equilibrium positions are unchanged from \eref{eq:cubic_z0}. The normal-mode frequencies, however, are now
\begin{equation}
	\omega_\pm \simeq \sqrt{\frac{2 q \kappa_2}{m_1}} \left(1+\mu\pm\sqrt{\mu^2-\mu+1}\right)^{1/2}
		\left(1\mp \frac{3}{2^{8/3}} \frac{1-\mu}{\sqrt{\mu^2-\mu+1}}\frac{l}{\lambda_3} \right) .
	\label{eq:cubic_unequal_mass}
\end{equation}
Note that the frequency shift is now first-order in $l/\lambda_3$ and depends on the ion order due to the first-order dependence on the sign of $l/\lambda_3$. In the limit of a purely harmonic potential, \eref{eq:cubic_unequal_mass} reproduces the results of \cite{01Morigi,00Kielpinski}. The normal-mode eigenvectors are
\begin{equation}
	e_{\pm}^{\prime} \simeq \frac{1}{\sqrt{1+r_\pm^2}}
	\left(\begin{array}{c}
		r_\pm\left[ \mp1 -\frac{3}{2^{5/3}} \frac{1+\mu}{\sqrt{\mu^2-\mu+1}} \frac{1}{1+r_\pm^2} \frac{l}{\lambda_3}\right]\\
		 1\mp\frac{3}{2^{5/3}}\frac{1+\mu}{\sqrt{\mu^2-\mu+1}}\frac{r_\pm^2}{1+r_\pm^2}\frac{l}{\lambda_3}
	\end{array}\right).
\end{equation}
Here, $r_\pm = [ \pm(\mu-1)+\sqrt{\mu^2-\mu+1}]/\sqrt{\mu}$. As for the equal-mass case, the higher-frequency mode (subscript +) involves the ions moving out of phase and the other mode involves in-phase motion. As before, the mode-amplitude corrections scale as $l/\lambda_3$, though with mass-dependent coefficients.

\subsubsection{Example 4: quartic term with two ions of unequal mass.}

For unequal-mass ions in a potential with a quartic perturbation, the normal-mode frequencies are
\begin{eqnarray}
	\omega_\pm \simeq& \sqrt{\frac{2 q \kappa_2}{m_1}} \left(1+ \mu \pm \sqrt{\mu^2-\mu+1}\right)^{1/2} \nonumber \\
		&\times	\left[1 + \frac{1}{3\cdot 2^{4/3}} \frac{\mp(1+\mu) + 7\sqrt{\mu^2-\mu+1}}{\sqrt{\mu^2-\mu+1}} \left(\frac{l}{\lambda_4}\right)^2 \right] .
\end{eqnarray}
In this case, the frequency shift is second-order in $l/\lambda_4$, as for the equal mass ions. As one would expect from the symmetry of the potential, the frequencies do not depend on ion order. The eigenvectors are
\begin{equation}
	e_{\pm}^{\prime} \simeq \frac{1}{\sqrt{1+r_\pm^2}}
	\left(\begin{array}{c}
		r_\pm \left[\mp 1 -\frac{1}{2^{1/3}} \frac{1-\mu}{\sqrt{\mu^2-\mu+1}} \frac{1}{1+r_\pm^2} \left(\frac{l}{\lambda_4}\right)^2\right]\\
		 1\pm \frac{1}{2^{1/3}} \frac{1-\mu}{\sqrt{\mu^2-\mu+1}} \frac{r_\pm^2}{1+r_\pm^2} \left(\frac{l}{\lambda_4}\right)^2
	\end{array}\right).
\end{equation}
Unlike the equal-mass case, the different masses break the symmetry, and there are quartic corrections to the mode amplitudes.

\subsection{\label{sec:Expt} Experimental demonstrations}

Using various combinations of \Beplus and \Mgplus, we demonstrate normal-mode frequency and amplitude shifts arising from weak anharmonic perturbations. We also present a technique that uses the ion order of an unequal-mass pair to suppress odd-order anharmonicities.

\subsubsection{Ion trap and potential wells.}

\begin{figure}[tbh]
 \includegraphics[width=3in]{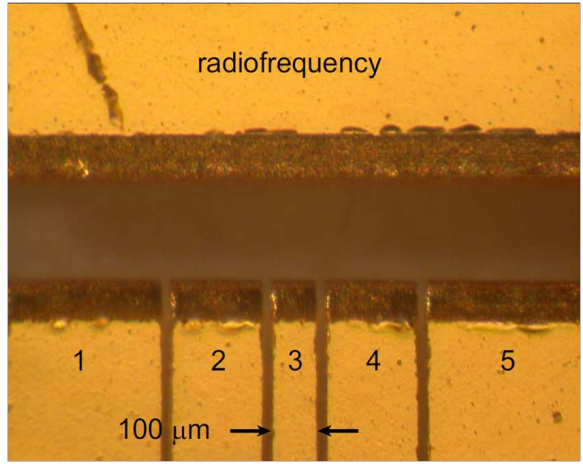} \centering \caption{A top view of one of the gold-plated alumina wafers that combines with a second underlying wafer to trap the ions. The picture encompasses the electrodes
in the region used in the experiments.}

\label{fig:trap}
\end{figure}

The linear radiofrequency Paul trap used here (described in \cite{09Jost}) stores the \Beplus and \Mgplus ions.
The trap is formed from two 125~$\mu$m thick wafers of alumina, with the electrodes made of sputtered gold and laser-cut 20~$\mu$m vacuum gaps between adjacent electrodes. A top view of the top wafer in
the relevant region of the trap is shown in \fref{fig:trap}.
Radial confinement of the ions is provided by a pseudopotential derived
from a quadrupole potential ($V_{\rm peak}\sim200-300$~V) oscillating at a radio frequency ($\Omega_\textrm{rf}=2\pi\times150$~MHz) and  resulting in radial secular
frequencies of approximately 12~MHz for a single \Beplus ion. The experimental
region of the trap has five pairs of control electrodes that can be used
to create axial potential wells (the trap has three additional pairs of
control electrodes, but these are placed far from the experimental
region and have only a small effect on the axial potentials in this
region). The pairs of electrodes are arranged such that one of each
pair is on the top wafer and the other opposes it on the bottom wafer \cite{02Rowe}. The nominal
voltage applied to the electrodes in one pair is the same; however
a (typically small) differential component is also used to null out
stray static electric fields at the pseudopotential zero as well as the effects of trap imperfections. In order
to design potential wells for the ions located at a particular position,
we simulate the influence of each electrode at that position using
the Boundary Element Method (BEM) \cite{BkPozrikidis}. Since the
fields from each electrode can be superposed, this allows us to design
sets of voltages that produce harmonic or anharmonic potential wells
at any point along the trap axis. It should be noted that while the
simulation provides guidance, it relies on dimensions
taken from photos of the actual trap, and thus its accuracy is limited.

\begin{figure}[tbh]
 \includegraphics[scale=0.7]{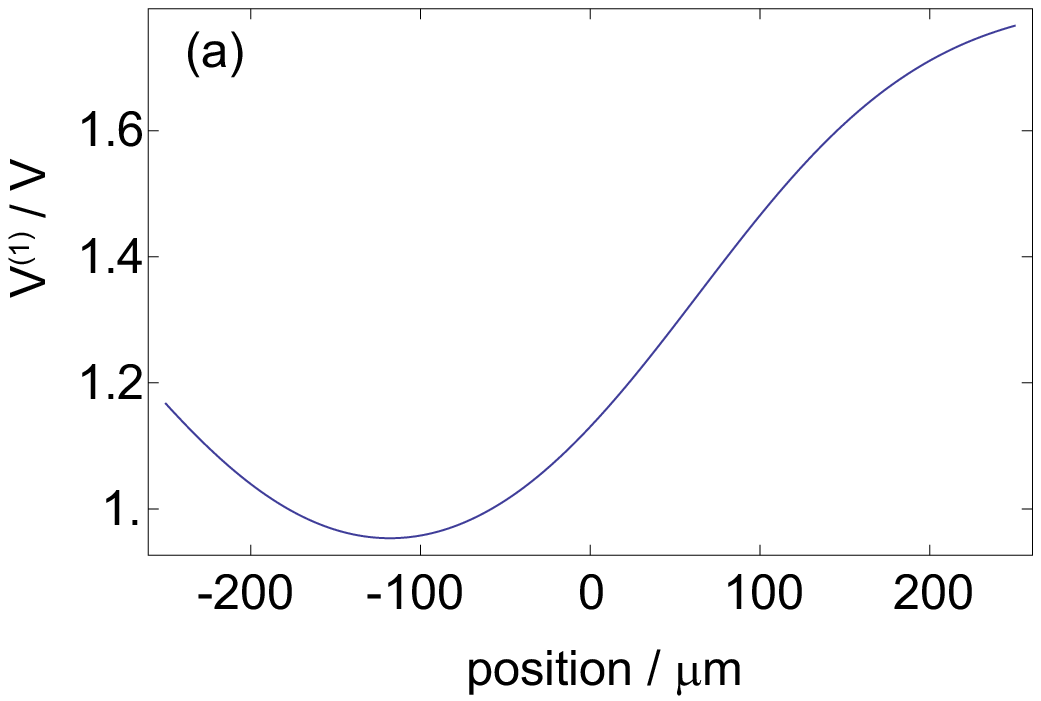} \includegraphics[scale=0.7]{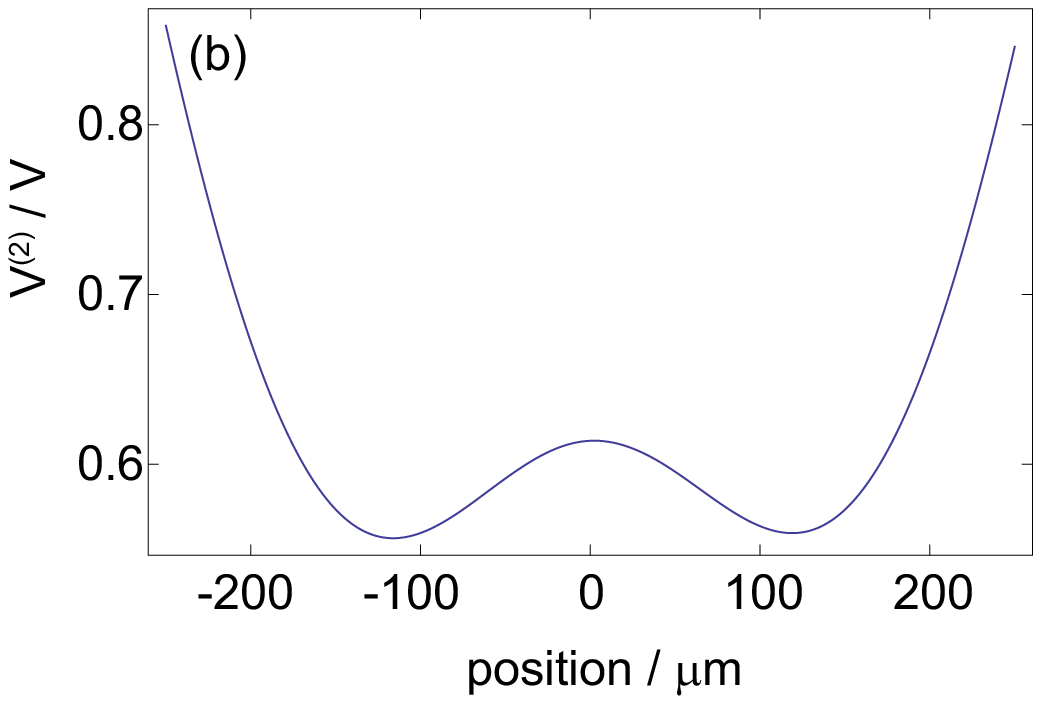}
\caption{Simulations of the two axial trapping potentials. The origin of the position co-ordinate is in the center of electrode 3.
(a) The harmonic-potential well, designed to have negligible anharmonic terms
along the axis of the trap. (b) The double-well potential, which has
significant anharmonicities in both wells. Zero corresponds to the centre of electrode 3.}

\label{fig:potential}
\end{figure}

For the quantum information experiments performed in this trap \cite{09Home,09Jost,10Hanneke},
we used two base sets of potentials. We periodically adjusted them with shim voltages or scaled them to adjust motional frequencies, but most experiments and waveforms began and ended with one of these sets. The first set produced a single potential
well, positioned close to electrode 2, with negligible
anharmonic terms along the axis of the trap. Despite our assumption above, slight electrode asymmetries can produce a small axial pseudopotential; therefore we designed this potential well to overlap its minimum position with the axial pseudopotential minimum, which we found experimentally.
The voltage configuration used to do this was $\bi{V}^{(1)}=\{3.7,1.3,1.5,4.1,3.7\}$~V,
where the numbers indicate the voltage applied to electrode pairs 1 to 5. The simulated potential
produced by these voltages is shown in \fref{fig:potential}(a). The quantum information experiments required simultaneous trapping of ions in two experiment zones, for which we used the double-well potential shown in \fref{fig:potential}(b). This potential used the voltage
set $\bi{V}^{(2)}=\{3.8,0.0,2.8,0.0,3.7\}$~V, producing two potential wells
240~$\mu$m apart. Both the single-well and the double-well potentials were designed to produce a minimum at the same position, -120~$\mu$m in \fref{fig:potential}, with the same curvature, $\kappa_2=m\omega_z^2/(2q)=1.3\times10^{7}~\textrm{V}\cdot\textrm{m}^{-2}$, corresponding to
$\omega_z=2\pi(2.7~\textrm{MHz})$ and $l = 3.8~\mu$m for beryllium ions. From BEM simulations
of the potential, we estimate the anharmonic terms of the double-well potential to be $\lambda_3 = -200~\mu$m 
and $\lambda_4 = 250~\mu$m. 
By comparison, simulations
of the single-well potential give $\lambda_3 = 10$~mm 
 and $\lambda_4 = 1$~mm. 

\subsubsection{Control and readout of trapped ions.}

Control of the ions' internal states and motion is primarily through standard techniques described elsewhere~\cite{09Jost,98Wineland2}. A 11.964~mT magnetic field establishes a quantization axis aligned at 45 degrees to both the trap axis and the vector normal to the alumina wafers.
The ions' motion is initially Doppler laser cooled to temperatures
low enough that the root-mean-square motional wavefunction size of
each mode is less than $100$~nm. For some experiments, we proceed with cooling one or more modes to near the ground state by use of Raman sideband transitions. The \Beplus ions' internal states are initialized
by optical pumping to the S$_{1/2}\ket{F=2,M_{F}=2}$ hyperfine level, which
we label as $|i\rangle$. To observe and Doppler cool the ions, we
use a 313~nm $\sigma^+$-polarized laser beam tuned close to resonance with the $\ket{i}\leftrightarrow$
P$_{3/2}$ $\ket{F^\prime=3,M_{F}^\prime=3}$ closed cycling transition. State-dependent resonant
fluorescence from the ions is detected on a photomultiplier tube
(PMT) during a 200~$\mu$s detection window. For an ion initialized
in $|i\rangle$, we observe an average of approximately eight counts on the PMT.

Transitions among internal and motional states of the \Beplus ions are induced by stimulated Raman interactions. These use two laser
beams derived from the same laser; the  pair of beams has a relative detuning equal to that of a hyperfine ``carrier'' transition or a hyperfine transition plus or minus a motional frequency (``sideband'' transitions). The laser is detuned -70~GHz from the
S$_{1/2}\leftrightarrow$ P$_{1/2}$ transition. The geometry of the beams relative to the ions determines whether the lasers interact with only the internal states or with the motion as well. For motion-sensitive transitions, both beams are
aligned at 45 degrees to the axis of the trap (one anti-parallel and one perpendicular to the magnetic field), with their difference
vector $\bdelta\bi{k}$ aligned along the axis, with a magnitude of $|\delta k|=2\pi\times\sqrt{2}/\lambda$,
where $\lambda=313$~nm.
We typically utilize transitions between the
states $\ket{\downarrow}\equiv$ S$_{1/2}\ket{F=2,M_{F}=1}$ and
$\ket{\uparrow}\equiv$ S$_{1/2}\ket{F=1,M_{F}=0}$, whose frequency difference has no first-order sensitivity to magnetic field~\cite{05Langer}. In order to make use of this transition,
we first transfer population from $|i\rangle\rightarrow\ket{\downarrow}$, again with a stimulated Raman interaction.
Subsequent to driving Raman transitions on $\ket{\downarrow}\leftrightarrow\ket{\uparrow}$,
we transfer $\ket{\downarrow}\rightarrow|i\rangle$ and $\ket{\uparrow} \rightarrow$ S$_{1/2}\ket{F=1,M_{F}=-1}$, where the latter has no transitions
that are close to resonance with the detection laser frequency and thus gives negligible
fluorescence.

To measure trap frequencies, we resonantly excite
the motion of the ions. An oscillating {}``tickle'' voltage is
applied to one electrode of the trap, resulting in an oscillating
electric field at the position of the ions. This excites motion when
tuned close to resonance with a normal mode. Large motional excitations $(\bar{n} > 100)$ may be observed as a decrease in fluorescence of state $|i\rangle$, but motion-sensitive stimulated Raman transitions enable the detection of motional excitation corresponding to a single quantum.
 Each method will be described in detail below.

\subsection{\label{sec:freqmeas} Frequency shifts, homogeneous ion crystals }

One way to characterize the anharmonicities is to examine the axial frequencies of the in-phase motion of chains of \Beplus ions as a function of the number of ions in the chain. With the ions
initialized in the internal state $|i\rangle$, we apply the tickle
voltage for 500~$\mu$s and subsequently
detect. A resonant force significantly
excites the motion, resulting in a drop in the ion fluorescence
level~(see for example \cite{wesenbergPRA2007}) as the ions' Doppler shifts approach the linewidth of the optical transition (for the fluorescence to drop to one half requires an excitation
of approximately 1000 phonons for typical trap frequencies). For each frequency
setting of the tickle voltage, we repeat the experiment 500 times,
and record the average number of photon counts observed during the 200~$\mu$s
detection period. We fit the results with a Lorentzian to obtain the
normal-mode frequency.

\begin{figure}[tbh]
\includegraphics[scale=0.7]{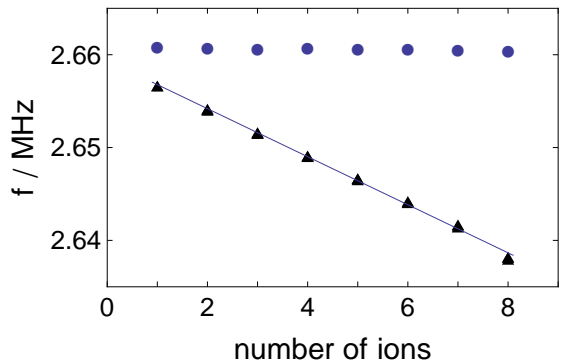} \centering \caption{The frequency of the centre-of-mass mode as a function of
the number of $^{9}$Be$^{+}$ ions. Triangles indicate the frequencies measured in the anharmonic
trap given by potentials $\bi{V}^{(2)}$; circles indicate the frequencies in the harmonic
trap given by potentials $\bi{V}^{(1)}$. The line is a fit to the anharmonic trap's frequency shift. Error bars are smaller than the symbols.}

\label{fig:tickleshift}
\end{figure}

The resonant frequencies for the in-phase mode of motion of linear
chains of one to eight \Beplus ions in the potentials created by $\bi{V}^{(1)}$
and $\bi{V}^{(2)}$ are shown in figure \ref{fig:tickleshift}. For the
nearly harmonic well, the in-phase mode frequency changes by -0.4(1)~kHz
between one ion and eight ions, an effect that is negligible on the plotted
scale. However in the case of the more anharmonic potential well, there is a frequency shift of $-2.59(3)$~kHz/ion. Both the cubic and quartic terms produce a linear frequency
shift per ion up to eight ions, so we are unable to distinguish these
components in this type of experiment.

\subsection{Frequency shifts, inhomogeneous ion crystals\label{sub:Frequency-shifts,-inhomogeneous}}

To isolate only the odd-order components of the anharmonicity, we
make use of an asymmetric ion chain, here consisting of one
\Beplus and one \Mgplus ion. We deterministically order the ions as \BM or \MB (see below). The length of the pair is $L = 4.8~\mu$m.
Odd-order anharmonicities produce different frequency shifts for different
ion order, whereas even terms produce identical shifts.

The mass difference between the ions and the mass dependence of the radial pseudopotential allow deterministic initialization of the ion order \cite{10HumeThesis,09Hume,10Chou}. The re-ordering
proceeds by first applying a radial electric field under continuous Doppler cooling so the ions remain close to their equilibrium positions. This field displaces
the \Mgplus ion more than the \Beplus one due to the weaker pseudopotential
experienced by the heavier-mass ion. At a critical value of the electric
field the axis of the two-ion crystal is normal to the trap axis. At this point, a differential shim voltage between upper and lower electrodes (e.g. electrodes 2-upper and 4-lower raised relative to 2-lower and 4-upper) can twist the axis of the static potential, breaking its symmetry relative to the orientation of the two ions, and
pushing the \Mgplus to a different value of $z$ compared to the \Beplus ion. By subsequently reducing the field and removing the twist, the ions are left in the desired order.

After ordering the ions, the frequency of the in-phase normal mode
of the two ions is measured by use of the tickle method. In the case of
two ions of unequal mass, both axial modes of motion involve modulation
in the distance between the ions. This means that anharmonicity in
the Coulomb interaction can produce frequency shifts on both modes as a function
of motional excitation \cite{08Roos}. These shift the deduced motional frequency,
and can prevent a single frequency drive from exciting the ions to
energies high enough to reduce fluorescence.

To reduce the
excitation energy at which a signal can be observed, we can observe weak
motional excitation using Raman transitions. We first initialize both
axial modes of the ion chain close to the ground state of motion by use of
Raman sideband cooling \cite{95Monroe}. After applying the tickle,
we probe the motional excitation of the \Beplus ion by resonantly driving
the $\ket{\downarrow}\leftrightarrow\ket{\uparrow}$ carrier transition with
the motion-sensitive Raman beams. We choose the drive duration such that
an ion in the ground state would make a full transition from from
$\ket{\downarrow}\rightarrow\ket{\uparrow}$, which would leave a
subsequent detection ``dark''. If the motion of a normal mode is excited
to state $|n\rangle$, the transition rate is reduced by a factor
given by the matrix element $\bra{n}\exp({i\eta(\hat{a}+\hat{a}^{\dagger})})\ket{n}$,
where $\eta=\delta k\,\sigma$ is the Lamb-Dicke parameter~\cite{98Wineland2}. This results
in incomplete population transfer, resulting in the detection of fluorescence
in the subsequent detection window. For the in-phase mode of motion
of a \BM pair, $\eta=0.18$, and the Rabi frequency of the transition
is reduced to approximately half the ground-state rate for $n=17$. A more precise description involves averaging the transferred population over the Fock state distribution of a coherent state, however the current estimate is sufficient for our present purposes.

\begin{figure}[tbh]
 \includegraphics[scale=0.7]{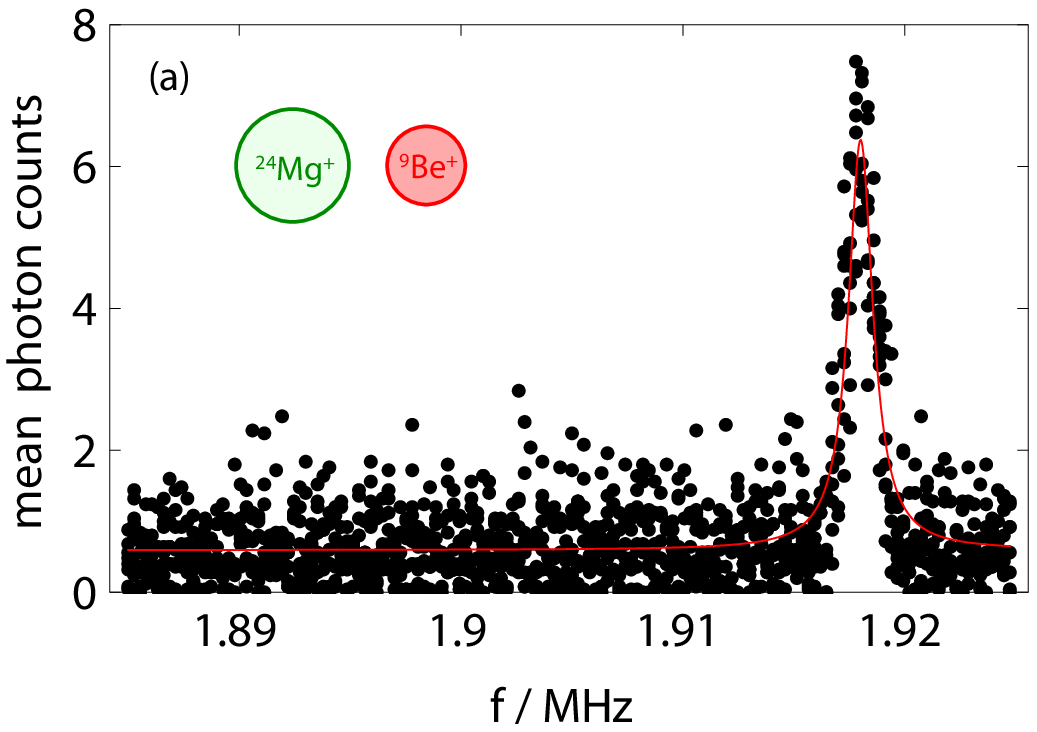} \includegraphics[scale=0.7]{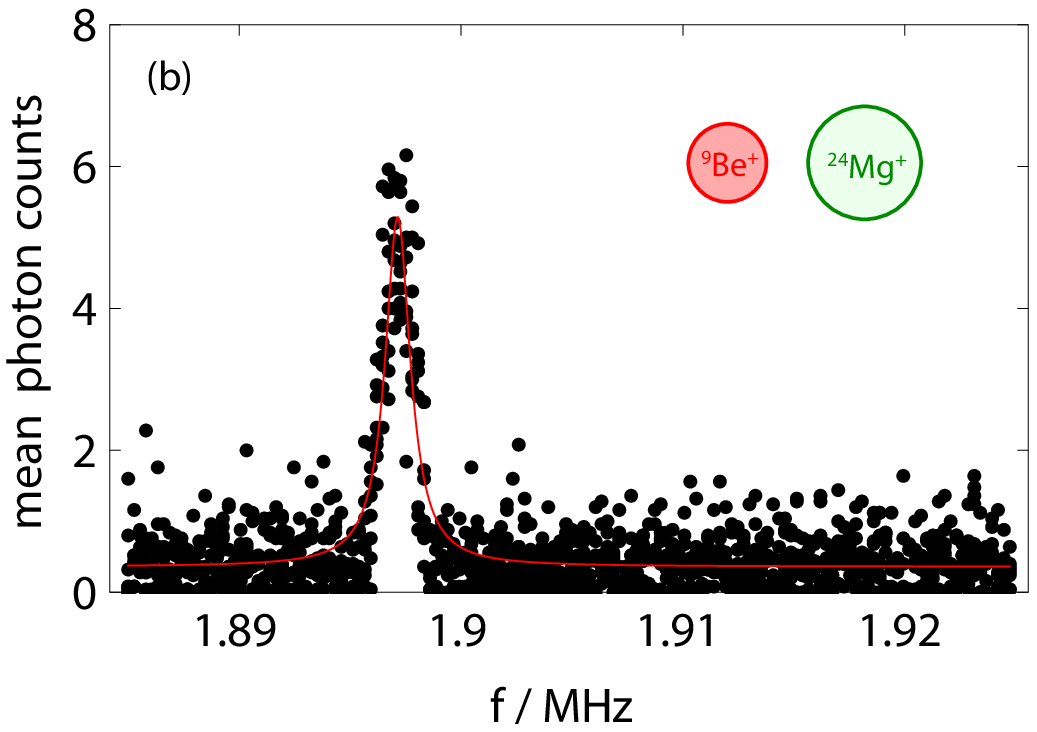}
\centering \caption{Frequency of the in-phase mode for ions in the order (a) \MB and (b)
\BM in the anharmonic well situated at -120~$\mu$m in figure \ref{fig:potential}.}

\label{fig:MgBeshift}
\end{figure}

Data from frequency scans of the tickle voltage applied near the in-phase axial mode frequency for two configurations
of ions in the anharmonic potential well characterized by $\bi{V}^{(2)}$ are shown in figure \ref{fig:MgBeshift}.
The data sets are each fitted with a Lorentzian, and the frequency
of the normal mode extracted. We measure a difference of 20.8(2)~kHz
between the mode frequencies for the two configurations. If the only
odd-order component is a cubic term, this corresponds to $\lambda_3 = -230~\mu$m. 
By comparison, a similar experiment performed in the more harmonic well characterized by $\bi{V}^{(1)}$
gives a frequency shift of less than 0.03(2)~kHz, corresponding to $|\lambda_3| > 150~$mm.

An even more sensitive method of observing motional excitation is
to probe on a motional sideband of the $\ket{\downarrow}\leftrightarrow\ket{\uparrow}$
transition. If the motional mode is in the ground state the motion-subtracting
sideband cannot be driven. However, when the motion is excited, population transfer is allowed between $\ket{\downarrow}$
and $\ket{\uparrow}$. To maximise the population in $n=1$ (for which
the Rabi frequency is set to make a full transfer), we would require
the tickle to excite the motion to a coherent state \cite{BkHaroche} with $\alpha\approx 1$. A coherent state
of this size would result in half the population being transferred
from $\ket{\downarrow}$ to $\ket{\uparrow}$. We use this method for nulling anharmonicity as described in section \ref{sec:null}.

\subsubsection{Measurement of modified ion amplitudes.}

When driving motion-sensitive Raman transitions, the amplitude of motion of
the ion in the normal mode affects the modulation index of the light,
thus affecting the Rabi frequency at which the mode is driven. Mathematically, for ion $j$,
the transition matrix element for the resonant transition between
$\ket{\downarrow_{j},n}$ and $\ket{\uparrow_{j},n+1}$ is proportional
to $\bra{n}\exp[i\delta k\,\sigma_{j}(\hat{a}+\hat{a}^{\dagger})]\ket{n+1}$~\cite{98Wineland2}, where
$\sigma_{j}$ is given in \eref{eq:rmsWavefunction} and is proportional to the ion's motional amplitude through the mode's eigenvector. For  $\sigma_{j}\ll 1/\delta k$,
this expression reduces to $\delta k\,\sigma_{j}\sqrt{n+1}$, showing direct proportionality to the amplitude.

In the experiment, we use a four-ion chain containing two \Beplus and
two \Mgplus ions, which are initialized in the order \BMMB prior to each run of the experimental sequence \cite{09Jost}. Only the beryllium ions interact with the 313~nm Raman light
fields. The length of the chain is $L = 10.8~\mu$m.

To equalize the intensity of each Raman light field on the two \Beplus ions, we separately observe the AC Stark shift of
each beam. Either beam alone cannot drive a transition between \ket{\downarrow} and \ket{\uparrow}, but the light causes a phase shift that we monitor with a Ramsey-type interference experiment~\cite{BkRamsey} where one beam is applied between the two Ramsey pulses. By equalizing the
rates at which these phases evolve for both ions, we equalize the relative
electric field strength of each light field at each ion to better than
2\,\%.

\begin{figure}[tb]
 \includegraphics[scale=0.7]{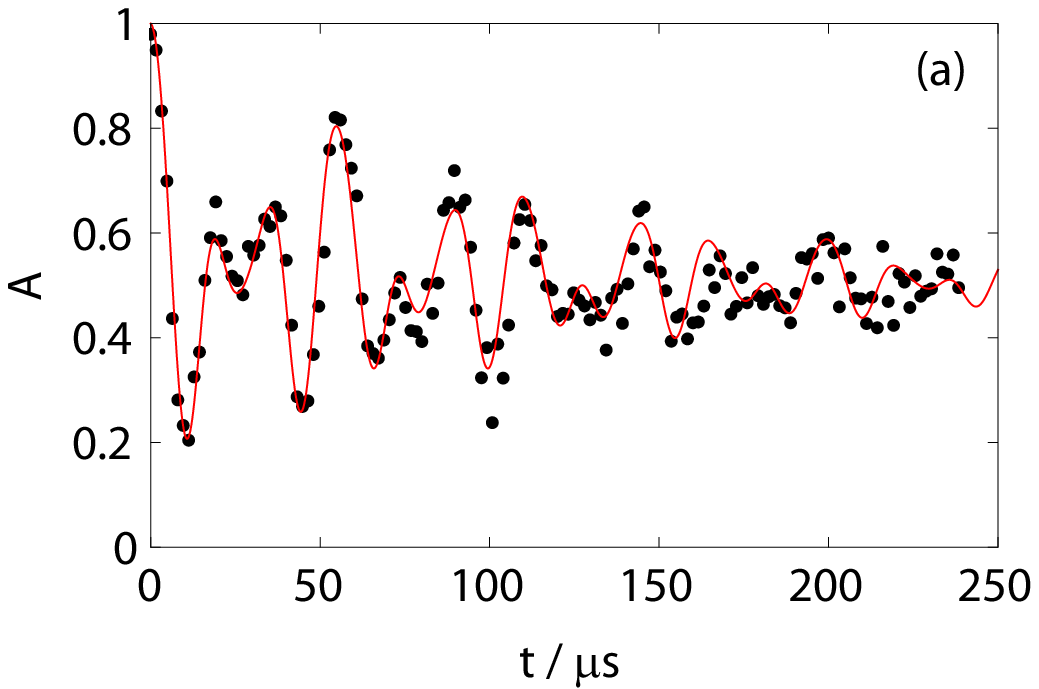} \includegraphics[scale=0.7]{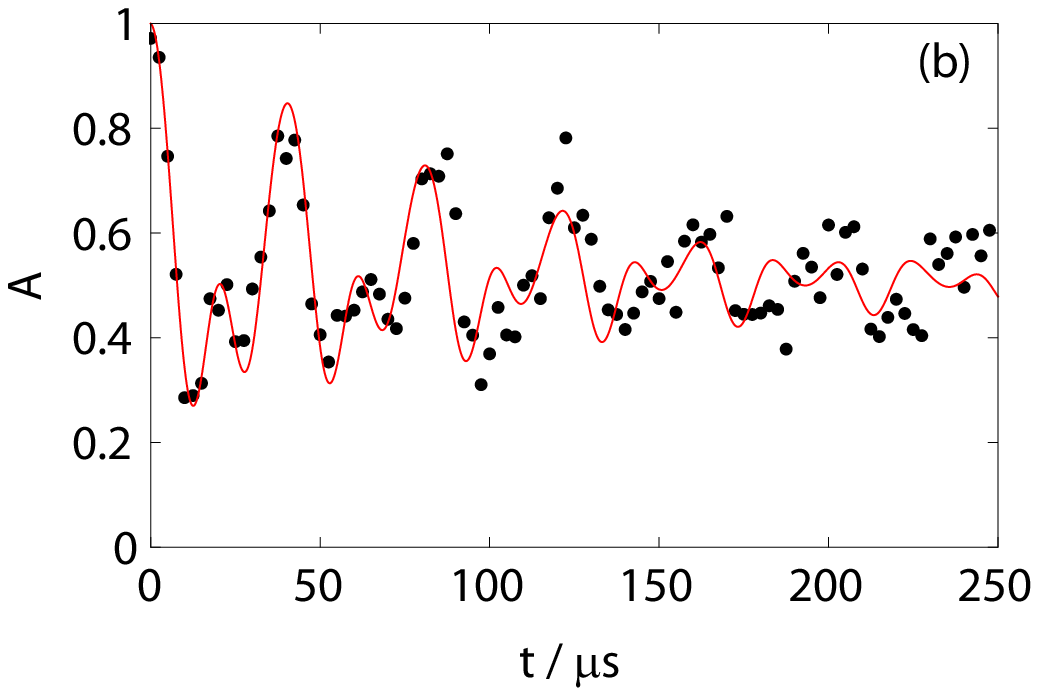}
\centering \caption{Experimental results and numerical simulations of sideband flopping
on the two Beryllium ions for (a) the third and (b) the fourth axial mode
of a 4-ion \BMMB chain in the anharmonic
trap. The amplitude is related to the population of ions in the $\ket{\uparrow}$ state, $A=P(\uparrow\uparrow)+[P(\uparrow\downarrow)+P(\downarrow\uparrow)]/2$,
and is proportional to the total fluorescence from both ions.}

\label{fig:sidebandflop}
\end{figure}

Once the light fields are equalized on the ions, we drive motional sidebands
of the four-ion chain. The two highest-frequency axial modes have frequencies of $f_3=5.5$~MHz and $f_4=5.7$~MHz and are particularly sensitive to the
cubic term in the axial potential. In the nearly harmonic potential they
have normal-mode eigenvectors $\bi{e}^\prime_{3}=(0.629,-0.322,-0.322,0.629)$
and $\bi{e}^\prime_{4}=(0.532,-0.465,0.465,-0.532)$. In the more anharmonic
potential, the amplitudes are significantly different. This becomes
obvious when the internal state populations are measured as a function
of sideband drive duration, where beating behavior is observed due to the different Rabi rates
of each ion. Results of such measurements are shown
in \fref{fig:sidebandflop}. Also shown are simulations of sideband
drives for two \Beplus ions with a ratio of amplitudes of ion motion ($R_\alpha = |e^\prime_{\alpha 1}/e^\prime_{\alpha 4}|$)
of $R_{3}=0.625$ and $R_{4}=0.500$
 (these values were chosen by eye to best fit the
data). The simulation curves are obtained by integrating the Schr\"odinger
equation, and subsequently adding a phenomenological decay to the Rabi oscillations in order to account for both motional and internal-state
decoherence. Both simulations use the same value of the carrier Rabi frequency.
For the cubic term of $\lambda_3 = -230~\mu$m 
obtained from the frequency
shifts of the \BM in-phase mode, we would expect these
ratios to be $R_{3}=0.644$ and $R_{4}=0.499$.
The eigenvectors in this case are $\bi{e}^\prime_{3}=(0.474,-0.167,-0.452,0.736)$
and $\bi{e}^\prime_{4}=(0.686,-0.531,0.359,-0.342)$. Though the agreement between the two values of $R_{3}$ seems good, choosing $R_{3}=0.644$ produces a noticeable mismatch between data
and theory. The reason for this discrepancy is currently not understood.

Unequal amplitudes of motion are generally undesirable for performing
multi-qubit logic gates on trapped ions. For the gate methods in common
use today \cite{08Blatt, 08Benhelm, 99Sorensen,03Leibfried}, driven ion motion during the gate adds a phase conditioned on the qubits' joint state. In these gates, the unequal mode amplitudes can lead to a longer required gate duration, thereby increasing the probability of error from off-resonant photon scattering \cite{07Ozeri}. Additionally, this phase could be different for \ket{\uparrow\downarrow} and \ket{\downarrow\uparrow} due to the different ion-amplitudes, which increases the complexity of gate calibration.

\subsection{\label{sec:null} Nulling odd-order anharmonicities}

For the purposes of optimizing quantum logic gates on multi-ion crystals,
it is desirable to be able to tune out the anharmonicities.
In the trap used here, we cannot do this while maintaining a double-well potential
if the positions of the minima of the two potential wells are fixed to be the same as the minima of the set of potentials $\bi{V}^{(2)}$ (the number of independent electrode voltages is insufficient to
satisfy all the required constraints). However, if we remove this position constraint, it is possible to produce two
wells with zero odd-order anharmonicity (though it seems that one voltage should not be sufficient to meet the constraints of the two wells, in practice the symmetry of the electrodes with respect to the well positions is sufficient to meet both constraints). We perform this optimization
by observing the ion-order dependence of the in-phase and out-of-phase mode frequencies of a \BM pair as a function
of the voltage $V_{3}$ applied to the control electrode positioned
between the two potential wells. The potentials used here are slightly modified from $\bi{V}^{(2)}$ and are optimized in simulation for negligible cubic anharmonicity in both wells. The nulling procedure allows in-situ adjustment to account for differences between simulation and experiment.

\begin{figure}[tb]
 \centering \includegraphics[scale=0.7]{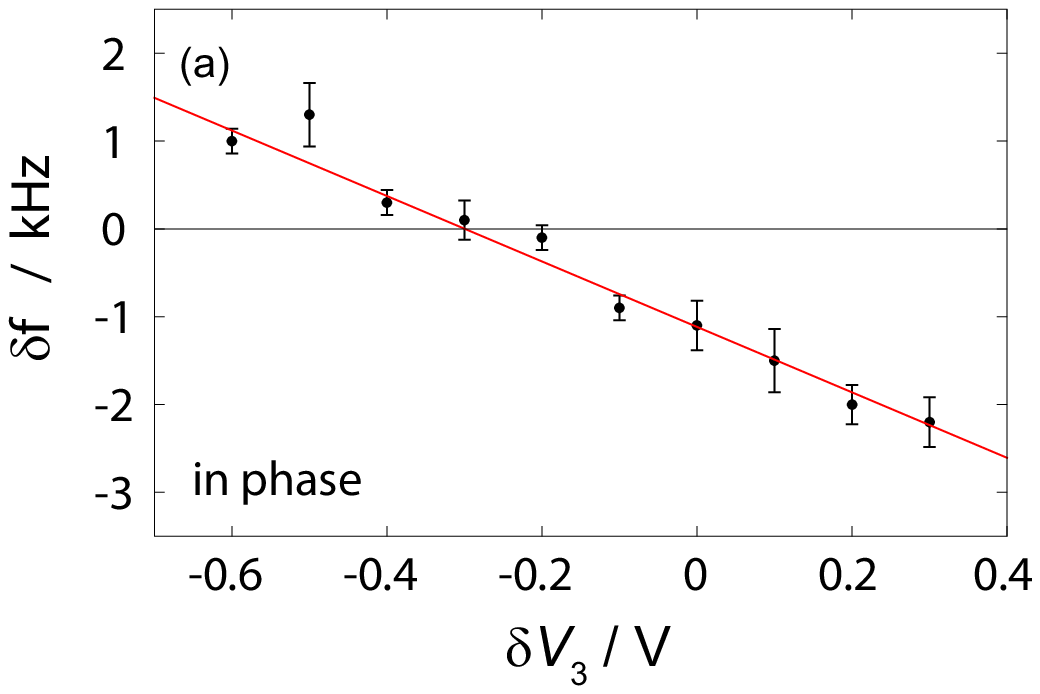} \includegraphics[scale=0.7]{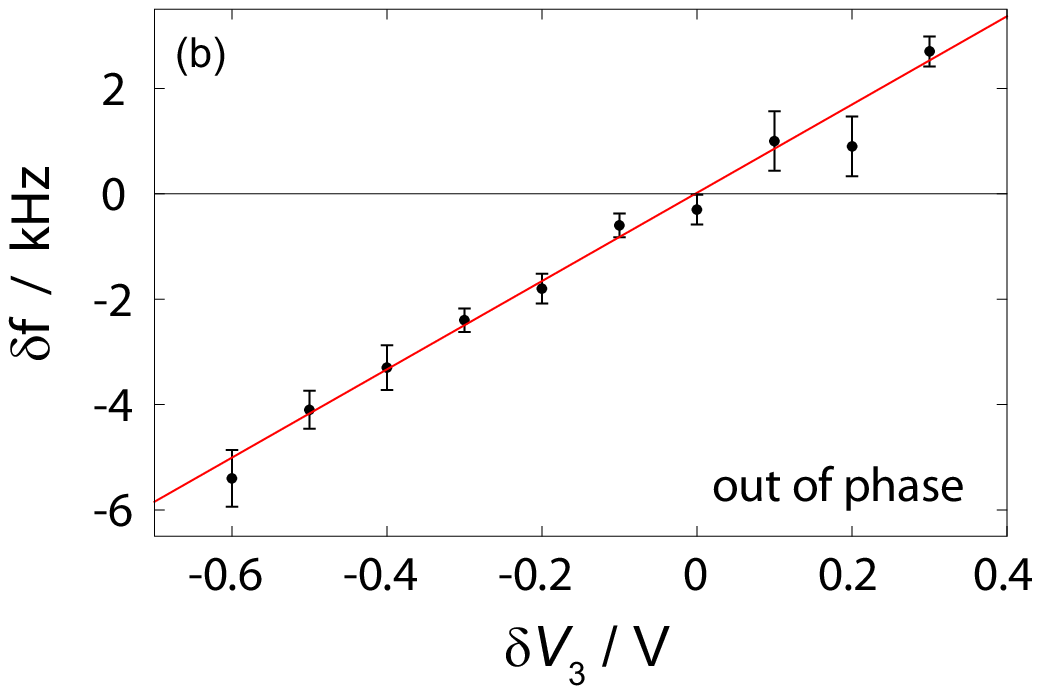}
\centering \caption{Ion-order-dependent frequency shift for the potential well close to electrode 2 in the double-well potential. The (a) in-phase and (b) out-of-phase mode frequency between the \MB
and \BM configurations as a function of electrode 3
voltage, which is plotted as an offset from the simulated value.}

\label{fig:CancelMgBeshift}
\end{figure}

Data from such an optimization are shown in \fref{fig:CancelMgBeshift}, which shows frequency differences when reversing ion order for the potential well close to electrode 2 in the double-well potential.
The data show that the value of $V_3$ where the in-phase mode shows zero frequency
shift on inverting the ion order is different from the value of $V_3$ where the
same condition is met for the out-of-phase mode. This discrepancy is consistent with a small pseudopotential gradient along the axis of the trap, indicating that the adjusted position is no longer at the axial pseudopotential minimum.
Since the pseudopotential has a different strength for both ions,
a gradient pointing from \Mgplus to \Beplus will
result in the spacing of the ions being reduced (and vice versa).
This tends to increase the energy of the out-of-phase mode, but has
a smaller impact on the in-phase mode. From the shift of the out-of-phase
mode at the point where the in-phase shift is nulled, we can calculate
a value of the pseudopotential gradient of $0.2$~eV$\cdot$m$^{-1}$.

To choose which value of $V_{3}$ to use to minimize
the \BMMB sideband amplitude imbalance
described in the previous section, we numerically calculated the amplitude
imbalance with the pseudopotential gradient fixed. We find that the
parameters for which the amplitudes are balanced are close to those
for which the \BM in-phase mode shift is minimized. For this condition, the resulting amplitudes
for the four ion chain are calculated to be $\bi{e}^\prime_{3}=(0.631,-0.323,-0.322,0.628)$
and $\bi{e}^\prime_{4}=(0.530,-0.465,0.467,-0.533)$.

\section{Strong anharmonic effects -- modifications to the normal-mode picture}

\label{sec:strong}

As anharmonic effects increase to the level where individual ions sample the anharmonicity during excursions about equilibrium (that is $\sigma/\lambda_n$ is no longer negligible), the independent normal-mode picture for ion motion is no longer valid. The anharmonicities lead to coupling between the normal modes.
The principal effects of this coupling are mode frequency
shifts as a function of the number of quanta in both the motional
mode of interest and other motional modes. Anharmonic
effects of this kind were observed in \cite{08Roos} where the mode
cross-coupling was due to anharmonicity in the Coulomb interaction
between two ions. A variety of couplings due to the Coulomb anharmonicity
are also discussed in \cite{03Marquet}. In what follows, we extend these results by considering shifts arising from an arbitrary applied trapping potential. As a concrete example, we
evaluate this in the context of a surface-electrode trap with the ion trapped $30~\mu$m above the electrode plane.

\subsection{Perturbation theory}

If the harmonic term still dominates the potential energy, we may treat anharmonicity as a perturbation. Our goal is to calculate the frequency shift $\Delta f_Z$ of a given mode (labeled $Z$ in this section) for the transition between $n_Z$ and $n_Z+1$. The shift may depend on the occupation of that mode $n_Z$ as well as that of all other modes $\{n_\alpha\}$. Because of the potential dependence on other modes, we again consider all three dimensions, as we did in section~\ref{sec:calc_normal_modes}, and revert to the notation used in that section. The cubic and quartic terms in a Taylor expansion of the potential energy $U$ will involve sums over the coefficients
\begin{eqnarray}
	A_{ijk}^{\prime(3)} & = &
		\left. \frac{1}{3!} \frac{1}{\sqrt{m_{i}m_{j}m_{k}}} \frac{\partial^{3}U}{\partial z_{i}\partial z_{j}\partial z_{k}}\right|_{\{z_{i}^{0}\}}\\
	A_{ijkl}^{\prime(4)} & = &
		\left.\frac{1}{4!} \frac{1}{\sqrt{m_{i}m_{j}m_{k}m_{l}}} \frac{\partial^{4}U}{\partial z_{i}\partial z_{j}\partial z_{k}\partial z_{l}}\right|_{\{z_{i}^{0}\}} ,
\end{eqnarray}
where the indices refer to ions not normal modes.
Both the Coulomb interaction and any anharmonic terms in the
trapping potential contribute to these coefficients.

Because we are interested in perturbations to the normal modes of the system, we transform these tensors into the normal-mode basis in a manner analogous to the vector case of \eref{eq:vector_transformation_into_normal_mode_basis}:
\begin{eqnarray}
	G_{\alpha\beta\gamma}^{\prime(3)} & = &
		\sigma^\prime_{\alpha} \sigma^\prime_{\beta} \sigma^\prime_{\gamma}
		\sum_{i,j,k}e_{\alpha}^{\prime i}e_{\beta}^{\prime j}e_{\gamma}^{\prime k}A_{ijk}^{\prime(3)}\\
	G_{\alpha\beta\gamma\delta}^{\prime(4)} & = &
		\sigma^\prime_{\alpha} \sigma^\prime_{\beta} \sigma^\prime_{\gamma} \sigma^\prime_\delta
		\sum_{i,j,k,l}e_{\alpha}^{\prime i}e_{\beta}^{\prime j}e_{\gamma}^{\prime k}e_{\delta}^{\prime l}A_{ijkl}^{\prime(4)}.
\end{eqnarray}
The inclusion of the $\sigma^\prime$ coefficients is for notational convenience. In this notation, the cubic and quartic terms of the Taylor expansion are
\begin{eqnarray}
	U^{(3)} = \sum_{\alpha,\beta,\gamma} G^{\prime (3)}_{\alpha\beta\gamma} (\hat{a}_\alpha+\hat{a}_\alpha^\dagger) (\hat{a}_\beta+\hat{a}_\beta^\dagger) (\hat{a}_\gamma+\hat{a}_\gamma^\dagger), \\
	U^{(4)} = \sum_{\alpha,\beta,\gamma,\delta} G^{\prime(4)}_{\alpha\beta\gamma\delta}
		(\hat{a}_\alpha+\hat{a}_\alpha^\dagger) (\hat{a}_\beta+\hat{a}_\beta^\dagger) (\hat{a}_\gamma+\hat{a}_\gamma^\dagger) (\hat{a}_\delta+\hat{a}_\delta^\dagger) .
\end{eqnarray}
Here, we have written the position operator $\hat{\zeta}^\prime_\alpha$ in terms of raising and lowering operators as in \eref{eq:quantized_position_operator}.

In first-order perturbation theory, all odd-order contributions vanish exactly and the leading contribution is from the quartic term:
\begin{equation}
	\Delta E_1 \left( n_\alpha, n_\beta, n_\gamma, n_\delta \right) =
		\left\langle n_\alpha, n_\beta, n_\gamma, n_\delta  \left|
			U^{(4)}
		\right| n_\alpha, n_\beta, n_\gamma, n_\delta  \right\rangle .
	\label{eq:first_order_perturbation}
\end{equation}
Only terms with coefficients $G^{\prime(4)}_{ZZZZ}$ and $G^{\prime(4)}_{\alpha\alpha ZZ}$ have nonzero contributions to the $n_Z \leftrightarrow n_Z+1$ frequency shift.

The cubic terms give the leading contribution in second-order perturbation theory,
\begin{equation}
	\fl  
	\Delta E_2 \left( n_\alpha, n_\beta, n_\gamma \right) =
	\sum_{\{\tilde{n}_\alpha, \tilde{n}_\beta, \tilde{n}_\gamma\} \atop \neq \{n_\alpha, n_\beta, n_\gamma\}}
	\frac{\left|\left\langle \tilde{n}_\alpha, \tilde{n}_\beta, \tilde{n}_\gamma \left|
		U^{(3)}
		\right| n_\alpha, n_\beta, n_\gamma \right\rangle\right|^2}
	{\hbar\left[\omega_\alpha\left(n_\alpha-\tilde{n}_\alpha\right)+
			\omega_\beta\left(n_\beta-\tilde{n}_\beta\right)+
			\omega_\gamma\left(n_\gamma-\tilde{n}_\gamma\right) \right]}.
	\label{eq:second_order_perturbation}
\end{equation}
These terms can potentially lead to a number of resonances. For example, it is possible for two modes to have a resonance allowing the destruction of two phonons from one of the modes and the creation of one phonon in the other. The $G^{\prime(3)}_{\alpha\alpha Z}$ term in the potential creates such an interaction. In this discussion, we assume we are detuned from all such resonances so that perturbation theory remains valid. As ion numbers increase, mode density will increase and it will be harder to avoid these resonances.

After evaluating the ladder-operator algebra for the leading perturbation-theory contributions above, we find that the frequency shift for the $n_{Z}\leftrightarrow n_{Z}+1$ transition is
\begin{eqnarray}
\fl  
h\Delta f_{Z}(\{n_{\alpha}\},n_{Z}) & = & \Delta E_1(\{n_{\alpha}\},n_{Z}+1)-\Delta E_1(\{n_{\alpha}\},n_{Z})\nonumber \\
 && + \Delta E_2(\{n_{\alpha}\},n_{Z}+1)-\Delta E_2(\{n_{\alpha}\},n_{Z})\nonumber \\
& = & 12 \left[\left(n_Z+1\right)G_{ZZZZ}^{\prime(4)} + \sum_{\alpha\neq Z}^{3N} G_{\alpha\alpha ZZ}^{\prime(4)}\left(1+2n_\alpha\right) \right] \nonumber \\
&&-\frac{36}{\hbar}\sum_{\alpha\neq Z}^{3N}(2n_{\alpha}+1)\left[\frac{2\omega_{\alpha}(G_{\alpha\alpha Z}^{\prime(3)})^{2}}{4\omega_{\alpha}^{2}-\omega_{Z}^{2}}+\frac{2\omega_{Z}(G_{ZZ\alpha}^{\prime(3)})^{2}}{4\omega_{Z}^{2}-\omega_{\alpha}^{2}}\right.\nonumber \\
 &&\quad + \left.\frac{G_{ZZZ}^{\prime(3)}G_{\alpha\alpha Z}^{\prime(3)}}{\omega_{Z}}+\frac{G_{\alpha ZZ}^{\prime(3)}G_{\alpha\alpha\alpha}^{\prime(3)}}{\omega_{\alpha}}\right]\nonumber \\
 && - \frac{6}{\hbar}(n_{Z}+1)\left[10\frac{(G_{ZZZ}^{\prime(3)})^{2}}{\omega_{Z}}-6\sum_{\alpha\neq Z}^{3N}\frac{(G_{ZZ\alpha}^{\prime(3)})^{2}\omega_{\alpha}}{4\omega_{Z}^{2}-\omega_{\alpha}^{2}}+12\sum_{\alpha\neq Z}^{3N}\frac{(G_{\alpha ZZ}^{\prime(3)})^{2}}{\omega_{\alpha}}\right]\nonumber \\
 && - \frac{72}{\hbar} \sum_{\alpha\neq Z}^{3N}\sum_{\beta\neq Z,\alpha}^{3N} (G_{\alpha\beta Z}^{\prime(3)})^{2} \left[  \frac{(n_{\alpha}-n_{\beta})(\omega_{\beta}-\omega_{\alpha})}{(\omega_{\beta}-\omega_{\alpha})^{2}-\omega_{Z}^{2}}+\frac{(n_{\alpha}+n_{\beta}+1)(\omega_{\beta}+\omega_{\alpha})}{(\omega_{\beta}+\omega_{\alpha})^{2}-\omega_{Z}^{2}} \right ] \nonumber \\
 && - \frac{36}{\hbar}\sum_{\alpha\neq Z}^{3N}\frac{G_{\alpha ZZ}^{\prime(3)}}{\omega_{\alpha}}\left[\sum_{\beta\neq Z,\alpha}^{3N}G_{\alpha\beta\beta}^{\prime(3)}(2n_{\beta}+1)\right].\label{eq:freqshift}\end{eqnarray}

\begin{figure}[bt]
	\centering \includegraphics[width=5in]{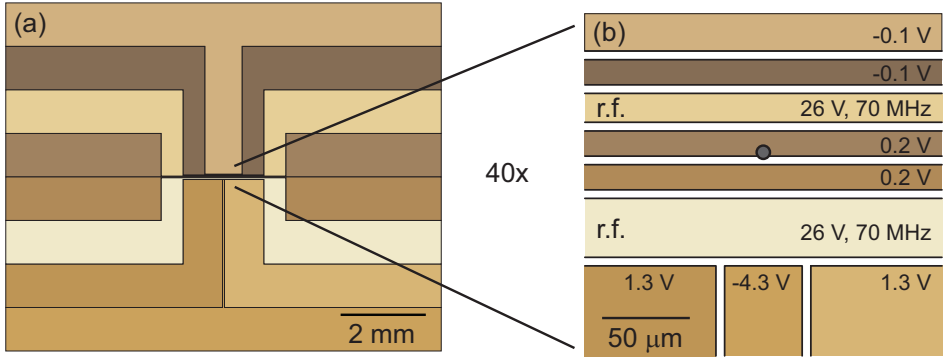}
\centering \caption{Diagram of (a) a microfabricated surface-electrode trap and (b) a closer view at the trapping region. Colors differentiate the electrodes, which are marked with their bias voltages. The radiofrequency electrodes are at DC ground. A single ion would be trapped near the circle in part (b) and 30~$\mu$m above the surface; two ions would lie approximately $L = 4.5~\mu$m apart with their axis parallel to the electrode axes. The intrinsic asymmetry of the electrodes relative to an ion's position leads to anharmonic perturbations in the trapping potential.}
\label{fig:SurfaceTrap}
\end{figure}

\subsection{Example case: a microfabricated surface trap}

One of the smaller traps that has been used at NIST has the electrodes lying in a plane and \MgplusH ions trapped
30~$\mu$m above that surface (a nearly identical trap was used in \cite{11Ospelkaus}). \Fref{fig:SurfaceTrap} shows the trap geometry. The
trap is operated with single-ion secular frequencies for the radial modes of
$f_1=7~\textrm{MHz}$ and $f_2=5~\textrm{MHz}$, and an axial frequency of $f_3=1.8~\textrm{MHz}$. The asymmetry
of the trap electrodes with respect to the position of the ion create cubic perturbations in directions out of the plane of the trap that dominate the
third-order tensor $G^{\prime(3)}$.

Using \eref{eq:freqshift}, we calculate a cross-coupling matrix $\chi$, which relates mode frequency shifts to excitation through
\begin{equation}
	\Delta f_Z = \sum_{\alpha=1}^{3N} \chi_{Z\alpha} \, n_\alpha .
	\label{eq:strong_anharmonic_freq_shift}
\end{equation}
For a single ion in this trap,
\begin{eqnarray}
		\left(\begin{array}{r}
		\Delta f_{1}\\
		\Delta f_{2}\\
		\Delta f_{3}\end{array}\right)
		=\left(\begin{array}{rrrrrr}
		-2.9 & -2.7 & 0.04\\
		-2.7 & -0.9 & 0.2\\
		0.04 & 0.2 & -0.1\end{array}\right)
		\left(\begin{array}{r}
		n_{1}\\
		n_{2}\\
		n_{3}\end{array}\right)
		{\rm Hz.}
	\label{eq:shift1ion}
\end{eqnarray}
The largest shifts are a few parts in $10^7$ per quantum.

Multi-qubit gates in quantum information processing typically involve multiple ions in the same trap zone. The couplings among the six normal modes of two trapped ions for the trap of \fref{fig:SurfaceTrap} are
\begin{equation}
	\left(\begin{array}{r}
	\Delta f_{1}\\
	\Delta f_{2}\\
	\Delta f_{3}\\
	\Delta f_{4}\\
	\Delta f_{5}\\
	\Delta f_{6}\end{array}\right)
	=\left(\begin{array}{rrrrrr}
	-1.4 & -3.2 & -1.3 & -1.6 & 0.03 & 0.03\\
	-3.2 & -0.4 & -2.2 & -2.1 & -9.4 & 0.03\\
	-1.3 & -2.2 & -0.4 & -1.1 & 0.2 & 0.1\\
	-1.6 & -2.1 & -1.1 & 1.6 & -13.5 & 0.3\\
	0.03 & -9.4 & 0.2 & -13.5 & 6.5 & -0.4\\
	0.03 & 0.03 & 0.1 & 0.3 & -0.4 & -0.1\end{array}\right)
	\left(\begin{array}{r}
	n_{1}\\
	n_{2}\\
	n_{3}\\
	n_{4}\\
	n_{5}\\
	n_{6}\end{array}\right)
	{\rm Hz.}
\end{equation}
The tensor is ordered such that the highest-frequency normal
mode is at the top (index 1), and the lowest is 6. Modes 5 and 6 are axial, and modes 1--4 are radial. The modes where the ions' amplitudes have opposite sign (the out-of-phase modes) are
2, 4 and 5, while the in-phase modes are 1, 3 and 6. For comparison,
if only the contribution from the Coulomb interaction is included (that is, if the trap potential were perfectly harmonic, having the same single ion trapping frequencies as above),
the shifts are
\begin{equation}
	\left(\begin{array}{r}
	\Delta f_{1}^c\\
	\Delta f_{2}^c\\
	\Delta f_{3}^c\\
	\Delta f_{4}^c\\
	\Delta f_{5}^c\\
	\Delta f_{6}^c\end{array}\right)
	=\left(\begin{array}{rrrrrr}
	0 & 0 & 0 & 0 & 0 & 0\\
	0 & 1.1 & 0 & 0 & -9.4 & 0\\
	0 & 0 & 0 & 0 & 0 & 0\\
	0 & 0 & 0 & 2.2 & -13.7 & 0.1\\
	0 & -9.4 & 0 & -13.7 & 6.7 & -0.1\\
	0 & 0 & 0 & 0.1 & -0.1 & 0\end{array}\right)
	\left(\begin{array}{r}
	n_{1}\\
	n_{2}\\
	n_{3}\\
	n_{4}\\
	n_{5}\\
	n_{6}\end{array}\right)
	{\rm Hz}.
\end{equation}
On comparing these matrices we observe that in this trap the
Coulomb anharmonicity still creates the largest mode-couplings, although some terms arising from the trap potential approach the strength of the Coulomb contribution. However whereas the Coulomb anharmonicity
principally affects only modes of motion where the oscillation of
the two ions has a sizeable differential component to the motion, the trap potential anharmonicities couple all modes.

\subsection{Coherence during quantum state manipulations}

\label{sec:coherence} The frequency shifts described in the previous
section can become problematic both for motional state engineering
\cite{96Meekhof,96Monroe,04Haljan,07McDonnell,10Zahringer,09Jost} and in quantum information processing,
where the motion of the ions is used in multi-qubit gates. To give an idea of the level at which cross-coupling from anharmonic terms could
impede control in typical experiments, we provide
two examples: the coherence of a superposition of motional Fock states and loss of fidelity in a two-qubit gate.

Consider the superposition
$(\ket{0_{Z}}+\ket{n_{Z}})/\sqrt{2}$ for a mode $Z$, and assume for simplicity that all other ``spectator''
modes can be described by a single Doppler-limit temperature $T_D$~\cite{leibfriedRMP2003}. The coherence between
the two motional states is given by the off-diagonal element of the density matrix, which evolves in time because of the anharmonic frequency differences. Taking the thermal average of the motional excitation of spectator modes yields
\begin{eqnarray}
	C(t) &=& 2 \left|\rho_{0_Z,n_Z}\right| =
			\left| \left\langle e^{ i 2\pi \left[f_Z(0)-f_Z(n_Z)\right]t } \right\rangle_{\!\{n_\alpha\}} \right| \nonumber \\
		&=& \left| \prod_{\alpha\neq Z}
			\left[1-\exp\left( \frac{-h f_\alpha}{k_B T_D}\right)\right]
			\left[1-\exp\left( \frac{-h f_\alpha}{k_B T_D}-i 2\pi \chi_{Z\alpha}\, n_Z t\right)\right]^{-1} \right| .
		\label{eq:Fockcoherence}
\end{eqnarray}
For example, if the superposition is between the ground and first excited states of the highest-frequency
radial mode (mode 1) of a single \MgplusH ion in the surface trap presented
in the previous section, and the other two modes are cooled to $T_D =  0.7~\textrm{mK}$, the coherence decays to 1/2 in 40~ms. Because the cross-couplings are themselves coherent, $C$ recovers to approximately 80~\% at $1/\chi_{1\,2}\approx 0.4~\textrm{s}$ and in principle almost fully recovers at $1/\chi_{1\,3}\approx25~\textrm{s}$. Such recoveries have been observed in other experiments~\cite{08Roos}. For a motional superposition with $(\ket{0} + \ket{10})/\sqrt{2}$, the decay is ten-times faster.
Recent experiments in other traps have produced superpositions
of motional states with occupations greater than $n=100$,
which were manipulated over time scales of milliseconds \cite{10Zahringer}.
If similar experiments were performed in the surface trap described here, the influence of the mode cross-coupling would need to be taken into consideration.

All deterministic multi-qubit gates in trapped-ion quantum information processing rely on transient excitation of motion conditioned on the ions' internal (qubit) states~\cite{99Sorensen,03Leibfried,11Ospelkaus,95Cirac,03GarciaRipoll}. At present, typical implementations involve cooling the ions to the ground state of one or more modes of motion prior to implementing the gate.
To increase the processing speed,
it would be advantageous to perform high-fidelity quantum logic
gates after only Doppler cooling. This is possible when the ions are well within the Lamb-Dicke regime and such gates have been demonstrated using lasers~\cite{kirchmairNJP2009} and magnetic-field gradients~\cite{11Ospelkaus}. In these cases, all the motional modes are thermally occupied with non-zero $\bar{n}$, and anharmonic effects need to be considered. The gate methods used in these demonstrations are different versions of a common type, which make use of transient motional excitation by an internal-state-dependent force. This force may be applied optically~\cite{99Sorensen,03Leibfried, 05Lee} or via microwaves~\cite{11Ospelkaus,08Ospelkaus}, where the form of the state-dependence can be tailored by control of the driving fields.

Here for simplicity, we consider the approach taken in \cite{03Leibfried}, which assumed a Hamiltonian
\be
H =  \hbar \Omega(t) \hat{S}_z \cos{(\omega t)} \eta (\hat{a} e^{-i \omega_Z t} + \hat{a}^\dagger e^{i \omega_Z t})
\ee
where $\Omega(t)$ is related to the laser fields used to produce the gate, $\omega$ is a drive frequency typically near $\omega_Z$, $\hat{S}_z$ is the sum of the Pauli $Z$ operators $\ket{\uparrow}\bra{\uparrow} - \ket{\downarrow}\bra{\downarrow}$ acting on each ion's spin, $\eta = |\delta k \sigma_1| = |\delta k \sigma_2|$ is the Lamb-Dicke parameter for the interaction between the light fields and the two ions involved in the gate, and $\hat{a}^\dagger$ and $\hat{a}$ are the creation and annihilation operators which act on the mode chosen for the gate. For simplicity, in what follows we assume that $\Omega(t) = \Omega$ for $0 < t < \tau_G$ and is zero otherwise. If $|\Omega|, |\omega - \omega_Z| \ll |\omega_Z|, |\omega|$ we can make a rotating wave approximation with respect to the motional frequencies, resulting in an evolution operator for the system given by
\be
\hat{U}(t) = D(\alpha(t) \hat{S}_z) \exp(i \Phi(t) \hat{S}_z^2)
\ee
where $D(\beta)$ is the motional state displacement operator $\exp(\beta \hat{a}^\dagger - \beta^* \hat{a})$ \cite{BkHaroche}, and
\be
\alpha(t) = -\frac{\Omega}{\delta} e^{-i \delta t/2} \sin(\delta t/2) ,\\
\Phi(t) = \frac{\Omega^2}{4 \delta^2} \left[\sin(\delta t) - \delta t \right],
\ee
with $\delta = \omega - \omega_Z$.

A common method for characterizing the performance of such a gate is to examine the fidelity with which the entangled state $\ket{\psi_{\rm ideal}} = \left(\ket{\downarrow  \downarrow}-i\ket{\uparrow \uparrow}\right)/\sqrt{2}$ is produced from the state $\ket{\downarrow \downarrow}$ when the force pulse is applied in the first half of a spin-echo sequence and $\Omega = \delta$ and $\tau_G = 2 \pi/\delta$. Ideally, these values result in $\alpha(\tau_G) = 0$ and $\Phi(\tau_G) = \pi/2$. For general $\alpha(\tau)$, $\Phi(\tau)$, the fidelity for producing $\ket{\psi_{\rm ideal}}$ at the end of the spin-echo sequence is given by \cite{03Leibfried}
\be
F = \left| \bra{\psi_{{\rm ideal}}} \rho \ket{\psi_{{\rm ideal}}} \right| = \frac{3}{8} + \frac{1}{8} e^{-2 |\alpha(\tau)|^2} + \frac{1}{2} e^{- |\alpha(\tau)|^2/2}\sin[\Phi(\tau)].
\ee

In the presence of anharmonicity, errors can enter into
this gate in three ways. First, if the motion begins in a distribution of states, each state will have a different mode frequency and therefore a different detuning. Thus, in general the motional state will not return to its initial position at the end of the drive. For a detuning $\delta(1 + \epsilon)$ with fractional error $\epsilon \ll 1$, this results in $\alpha(\tau_G) \simeq \pi \epsilon$. The resulting residual
entanglement between the internal states and the motion reduces fidelity when the motional degree of freedom is traced
out. Second, the different detunings cause each initial motional state
to enclose a different area of phase space and thus to acquire
a different phase. Again using a detuning error parameter $\epsilon$, we find that $\Phi(\tau_G) \simeq (\pi/2)(1 - 2 \epsilon)$. These phases must be averaged over using the distribution of initial motional states. Finally, the transient motional excitation itself will give a time-dependence to the detuning as the excitation increases then decreases during the gate. For the small excitations typical for present multi-qubit operations, this third effect is significantly smaller than the other two, and we ignore it in below.

Let us assume that the gate is performed with the duration chosen such that $\tau_G = 2 \pi/\delta$ for the detuning corresponding to all modes in the ground state. For ions which start in an incoherent motional state distribution with mean quantum numbers in the set $\{\bar{n}_\alpha\}$, the above expression then leads to a fidelity
\begin{equation}
	F =
	1-\frac{3 \pi^4}{\delta^2}
	\left( \sum_{\alpha\neq\beta} \chi_{Z\alpha}\chi_{Z\beta}\bar{n}_{\alpha}\bar{n}_{\beta}
		 +\sum_{\alpha=1}^{3N} \chi_{Z\alpha}^2 \overline{n_\alpha^2} \right) .
\end{equation}
The bars indicate an average over the incoherent distribution of $n$. For a thermal distribution, the mean-square $n$ in the last term can be rewritten as $\overline{n_\alpha^2} = \bar{n}_{\alpha}(2\bar{n}_{\alpha}+1)$. Failure of the motional states to return to their original position contributes two-thirds of the infidelity, with the other third due to the distribution of accumulated phases.

Consider two ions in the surface trap described in the previous section, and a two-qubit gate that makes use of the radial rocking mode with frequency $f_2=6.8$~MHz (as used in \cite{11Ospelkaus}). Implementing such a gate with all modes cooled to the Doppler-limit temperature of $T_D = 0.7$~mK produces an infidelity of
$1-F=4\times10^{-2} [2\pi(1~\textrm{kHz})/\delta]^2$ 
if all other aspects of the gate are perfect. For multi-qubit gates performed in a similar manner on larger numbers of ions \cite{10Monz, 05Leibfried}, the increased number of motional modes will reduce the fidelity even further. Thus for detunings $\delta > 2\pi(20~\textrm{kHz})$,
the infidelity is below the level of $10^{-4}$, which is
often estimated as a requirement for fault-tolerant quantum information
processing. Since the cubic terms in the potential scale unfavorably with reduced trap size, anharmonicity
may become a significant source of error as traps become smaller. Various approaches may be used to implement a gate in a manner that suppresses these errors. For example, though the detuning in the above example is $\delta = \Omega$, by use of two pulses---one performed in each half of the spin-echo sequence $\cite{06Home}$---the detuning could be increased to $\delta = \sqrt{2} \Omega$, thus reducing the error. This approach also has the advantage that the residual displacement arising from errors in detuning can be arranged to be of opposite sign for the two pulses, and hence is canceled out. This is a special case of the more general composite-pulse schemes recently suggested by Hayes et al. \cite{11Hayes}, that has been used in gate operations in several experiments \cite{ 09Home, 09Jost,  10Hanneke, 05Leibfried, 06Home}. Pulses with non-square shapes may also help in this regard~\cite{leibfriedPRA2007}. This method can be extended by increasing the detuning and executing more, but smaller, loops in phase space \cite{00Sorensen1}.  At ion--electrode distances of approximately $30$~$\mu$m, the coherence of
motional-state superpositions is currently limited by anomalous heating~\cite{00Turchette,06Deslauriers,08Labaziewicz},
which is larger than the anharmonic infidelity.

\section{Sensitivity to electric fields \label{sec:discussion}}

The range of effects described above focus on problems related
to anharmonic trapping potentials. In addition to these, a practical
consideration is that anharmonicities introduce a dependence of the
secular frequencies on the position of the ion. This means that uniform
electric fields can displace the ions and cause the trap frequencies to shift. For the two-layer trap described in \sref{sec:Expt} and a cubic
term with $\lambda_3 = -230~\mu$m 
(as observed in section \ref{sec:freqmeas}),
a field of 2~V$\cdot$m$^{-1}$ causes a fractional frequency shift of $\simeq10^{-3}$
(corresponding to 3~kHz in that example). For multiple experiments~\cite{09Home,09Jost,10Hanneke} performed using this trap, we observed that the trap frequency stability of the harmonic trapping potential
was better than that of the anharmonic potential. This suggests that frequency instability may have been caused by slow fluctuations in stray electric fields, but this correlation has not been characterized.

\section{Conclusion.}

Anharmonic trapping potentials give rise to a number of effects that
should be taken into account when performing quantum state engineering tasks in ion traps. As trap sizes are reduced,
these effects become more significant. By building traps with intrinsic geometric
symmetry, it should be possible to minimize odd-order anharmonicities. This
consideration should be made at the trap design stage. In practice,
the fabricated trap is unlikely to be an exact realization of any
design. For this purpose, it is desirable to include
a sufficient number of independently controllable electrodes to allow
anharmonicities to be canceled in situ. In this case, methods such
as those given in section \ref{sec:null} can be used to null out
unwanted terms. In addition to undesirable effects due to strong anharmonic trap potentials,
the advent of small traps where these terms can be engineered might
also enable novel state-preparation schemes \cite{03Marquet} and tunable phase transitions~\cite{08Fishman,gongPRL2010}.

\ack 

This work was supported by IARPA, NSA, DARPA, ONR, and the NIST Quantum Information Program. We thank Christian Ospelkaus for details of the surface-electrode trap and the use of \fref{fig:SurfaceTrap}. We thank John Gaebler and Ulrich Warring for comments on the manuscript. This manuscript is a contribution by the National Institute of Standards and Technology and is not subject to US copyright.

\section*{References}

\bibliography{myrefs}

\end{document}